# Storing magnetic information in IrMn/MgO/Ta tunnel junctions via field-cooling

D. Petti[1], E. Albisetti[1], H. Reichlová[2,3], J. Gazquez[4], M. Varela[5], M. Molina-Ruiz[6], A. F. Lopeandía[6], K. Olejník[2], V. Novák[2], I. Fina[4], B. Dkhil[7], J. Hayakawa[8], X. Marti[2,3,9], J. Wunderlich[10], T. Jungwirth[11] and R. Bertacco[1]

[1]LNESS-Dipartimento di Fisica, Politecnico di Milano, Via Anzani 42 Como, 22100, Italy

[2]Institute of Physics ASCR, v.v.i., Cukrovarnická 10 Praha 6, 162 53, Czech Republic

[3]Faculty of Mathematics and Physics, Charles University in Prague, Ke Karlovu 3 Prague 2, 121 16, Czech Republic

[4]Institut de Ciencia de Materials de Barcelona, ICMAB-CSIC, Universitat Autònoma de Barcelona, Bellaterra, 08193, Spain

[5]Materials Science & Technology Division, Oak Ridge National Laboratory, TN 37830 (USA) & Universidad Complutense de Madrid, 28040, Spain

[6]Departament de Fisica, Universitat Autonoma de Barcelona, Bellaterra, 08193, Spain

[7]Laboratoire Structures, Propriétés et Modélisation des Solides, UMR 8580 CNRS-Ecole Centrale Paris, 92295, France

[8]Hitachi Ltd., Advanced Research Laboratory, 1-280 Higashi-koigakubo, Kokubunju-shi, Tokyo, 185-8601, Japan

[9]Department of Physics, University of California, Berkeley, CA 94720, USA

[10]Hitachi Cambridge Laboratory, Cambridge, CB3 0HE, United Kingdom

[11]School of Physics and Astronomy, University of Nottingham, Nottingham, NG7 2RD, United Kingdom

**Abstract**

Tunneling junctions containing no ferromagnetic elements have been fabricated and we show that distinct resistance states can be set by field cooling the devices from above the Néel along different orientations. Variations of the resistance up to 10% are found upon field cooling in applied fields of 2T, in-plane or out of plane. Below $T_N$, we found that the metastable states are insensitive to magnetic fields thus constituting a memory element robust against external magnetic fields. Our work provides the demonstration of an electrically readable magnetic memory device, which contains no ferromagnetic elements and stores the information in an antiferromagnetic active layer.

Magnetic tunnel junctions used in modern hard-drive read heads and magnetic random access memories comprise two ferromagnetic electrodes whose relative magnetization orientations can be switched between parallel and antiparallel configurations, yielding the tunneling magnetoresistance effect.[1] Recently, large magnetoresistance signals have been observed on NiFe/IrMn/MgO/Pt stacks with an antiferromagnet (AFM) on one side and a non-magnetic metal on the other side of the tunnel barrier.[2,3] In these devices, ferromagnetic moments in NiFe are reversed by external magnetic field and the exchange-spring effect of NiFe on IrMn induces the rotation of the AFM moments in IrMn. This is then electrically detected via the measurement of the AFM tunneling anisotropic magnetoresistance (TAMR). The work has experimentally demonstrated the feasibility of a spintronic concept[4,5] in which the device transport characteristics are governed by an AFM.

The lack of magnetic stray fields and the relative insensitivity to external magnetic fields make AFM materials potentially fruitful complements to ferromagnets in the design of spintronic devices. The zero net moment of compensated AFMs, however, also implies that weak magnetic fields of the order of the typical magnetic anisotropy fields in magnets cannot be directly applied to rotate the AFM moments. In the devices reported in Refs. 2, 3 the problem was circumvented by attaching an exchange-coupled ferromagnet to the AFM electrode to form an exchange-spring.[6] This method, however, limits the thickness of the AFM layer to values not exceeding the domain wall width in the AFM. Since the exchange spring triggers rotation of the AFM moments at the opposite interface to the AFM/tunnel-barrier interface, the AFM TAMR effect can be observed only in AFM films which are thinner than the domain wall width in the AFM. Recent experiments in [Pt/Co]/IrMn/AlO$_x$/Pt stacks[7] have demonstrated that room-temperature AFM TAMR can be achieved in exchange-spring tunnel junctions only in a narrow window of AFM thickness. A subtle balance is required between a thin enough AFM to allow for the exchange-spring rotation of AFM moments across the entire width of the AFM and a thick enough AFM to avoid the decrease of the Néel temperature $T_N$ below room temperature by the size effects. We also point out that as a memory element, the exchange-spring AFM tunnel junctions can be disturbed by weak magnetic field perturbations as they still contain a ferromagnetic element.

To fully exploit the potential robustness of the AFM based spintronic device against magnetic fields, we have fabricated magnetic tunnel junctions analogous to those in Ref. 2,3, but without the auxiliary ferromagnetic NiFe layer. In these antiferromagnetic tunnel junctions (ATJs) we show that metastable states can be set by cooling the sample and crossing the Néel temperature in external magnetic fields with different orientations. These metastable states can be detected electrically, due to an analogous effect to the AFM TAMR reported in Refs. 2,3,7. Since our field-cooling approach for writing does not require any ferromagnetic layer, the limitation on the AFM thickness is removed in our devices. Our work provides the demonstration of an electrically readable magnetic memory device which contains no ferromagnetic elements and which stores the information in an AFM.

The stacks for the fabrication of the tunneling junctions used in this study have been deposited by magnetron sputtering (AJA ATC Orion 8 system). A Ta(20)/Ru(18)/Ta(2)/Ir0.2Mn0.8(2-8)/MgO(2.5)/Ta(20) (layer thicknesses are in nm) stack was deposited on SrTiO$_3$ (STO) single crystal after a chemical cleaning of the substrate. STO was chosen as a suitable insulating substrate with limited impact on the device behavior because our measurements are performed above 100K, i.e. in a temperature range where STO does not present structural transitions. The metallic layers have been deposited in dc mode, while MgO in rf mode. To reproduce the experimental conditions optimized for the fabrication of MgO based tunneling junctions, a magnetic field of 30 mT was applied along the STO[100] direction during the stack growth. A post growth annealing of 250°C has been performed in a dedicated system with an external magnetic field of 400 mT applied along the same STO[100] direction and with a field cooling until reaching room temperature (without crossing the Néel Temperature of IrMn).

The core of the stack employed for the fabrication of our devices is the $Ir_{0.2}Mn_{0.8}$(2-8)/MgO(2.5)/Ta(20) ATJ where the IrMn layer is in direct contact with the insulating barrier. In this way, modifications of its relativistic spin-orbit coupled band structure for different AFM configurations can yield the TAMR.[4] The high structural quality of our heterostructures is evident from Fig. 1, where we show STEM images taken on a sample with a 8 nm thick IrMn film. A Nion UltraSTEM operated at 100 kV and equipped with a Nion aberration corrector was used. Low and high resolution STEM Z contrast images show that the stacking comprises continuous films over large distances. This is consistent with atomic force microscopy analyses at intermediate growth steps, which revealed that each new layer preserved an RMS roughness of less than 1 nm (data not shown). The image shown in the inset of Fig. 1 highlights the successful recrystallization of MgO after annealing at 250°C for 1 hour. The MgO insulating barrier is highly textured along the out-of-plane [001] direction, parallel to the IrMn [111] texturation, as found in our preceding works[2] and confirmed by X-ray diffraction measurements (data not shown).

Pillar structures with different cross sectional areas, ranging from 4 to 100 $\mu m^2$, were patterned by optical lithography in order to define the ATJs for electrical measurements. The devices show tunneling I(V) characteristics and resistance area products (RA) typical of standard MgO magnetic tunneling junctions with the same MgO thickness (RA ~ $2.5 \cdot 10^5$ $\Omega$ $\mu m^2$ at 100 mV and 300 K). A Quantum Design Physical Property Measurement System (PPMS) and an Oxford Instruments cryostat furnished with vector magnet were used to perform the magneto-transport measurements.

As calorimetry measurements showed that $T_N$ of a 2 nm thick IrMn is reduced to ~173 K (see Fig. 3 and discussion thereafter), we performed the field cooling procedure from room temperature down to 120 K, thus covering a sufficiently large temperature range around $T_N$. In the case of data reported in the main panel of Fig. 2, we applied an external field $\mu_0 H_z$ of ±2T oriented perpendicular to the sample surface (scattered and red curves) and $\mu_0H_x = +2T$ oriented in the plane of the sample, along the [100] axis of the STO substrate (thick green curve). The RA product (measured at a fixed bias of 20 mV) is identical for the three field orientations in the temperature interval from 300 K to approximately 170 K, while below this temperature the RA traces for out-of-plane and in-plane field orientations split. At 120 K the difference is more than 10%. Remarkably, these states are metastable at temperatures sufficiently below $T_N$, as illustrated in the inset of Fig. 2. In the mesurement, the higher resistance state was prepared by in-plane field cooling and then the temperature was stabilized at 120 K. The RA product (at 20 mV bias) was then monitored while continuously sweeping the magnetic field in the out-of-plane direction ($H_z$) and also in the two orthogonal in-plane directions ($H_x$ and $H_y$) between +2 T and -2 T, for 10 hours. No changes in the tunneling resistance are observed within the experimental noise, which is much smaller than the difference between the higher and lower resistance states observed in the main panel of Fig. 2. This demonstrates that the state prepared by field cooling is metastable and insensitive to relatively large external magnetic fields. Note that metastable states showing different tunneling resistances at zero magnetic field were also observed in Ref. 2, where the configuration of the AFM moments in IrMn was controlled below $T_N$ using the exchange-spring effect of a ferromagnet. Noteworthy, the observed field-cool induced magnetotransport effect shows the key signatures of an anisotropic magnetoresistance. In Fig. 2, while the temperature-dependent resistance traces for the in-plane and out-of-plane fields split below $T_N$, we observe no difference between field-cool measurements performed at fields with the opposite polarity.

In Fig. 3 we highlight that the onset of the splitting of the RA traces for cooling in fields with different directions coincides with the transition to the ordered AFM state in the 2 nm IrMn film. Side by side we plot in the figure the normalized variation of the tunneling resistance $(R(H_x)-R(H_z))/R(H_z)$ in one of our 2 nm IrMn pillar devices and the differential specific heat of a 2nm IrMn layer as a function of temperature. Quasi-adiabatic nanocalorimetry (QAnC) is an ideal technique for investigating the Néel temperature of thin IrMn films. This technique allows for a direct measurement of the specific heat of the sample, enabling the observation of the critical behavior in the specific heat near $T_N$. Identical multilayers to those used in the tunnel junctions were sputtered onto self-standing silicon nitride membranes that form the nanocalorimetric cells. A twin calorimeter loaded with a reference multilayer sample without IrMn was used for differential measurements of the specific heat.[8] The Néel temperature inferred from the inflexion point of the specific heat singularity[8] is approximately 173 K, i.e., it is significantly reduced in the 2 nm IrMn film as compared to the bulk IrMn, having $T_N$ > 1000 K. This is in agreement with previous observations in case of other AFMs, e.g. CoO.[8] The reproducibility of the specific heat method has been confirmed in different samples prepared in separate growth runs under the same growth conditions. Moreover, the correspondence between $T_N$ and the onset of the field-cool AFM TAMR has been confirmed by independent measurements using the PPMS and the vector magnet cryostat, and studying different ATJs with the same nominal layer structure. All samples show a negligible magnetoresistance in the paramagnetic phase and a reproducible splitting of the RA traces when continuing the field-cooling below $T_N$ with in-plane and out-of-plane magnetic fields. The percentage difference between the two metastable resistance states obtained at 120 K varies from 2% to 10% in different ATJ samples. Higher values were found in devices with larger RA, thus indicating that tunneling is the origin of the observed magnetoresistance. The last one simply decreases in devices with thinner barriers, where defects can creates parallel conductive paths partially masking the effect of anisotropic tunneling.

We point out that the observed magnetoresistance cannot be ascribed to magnetization-independent tunneling transport phenomena due, e.g., to Lorentz force effects of the magnetic fields applied along different directions with respect to the tunneling current direction. These types of phenomena can be excluded since the field-cooling magnetoresistance disappears above $T_N$ and since we observe a negligible resistance variation upon application of external fields when the temperature is stabilized below $T_N$, as shown in the inset of Fig. 2. For the same reason, we exclude that possible Mn interdiffusion wthin the oxide layer is responsible for the observed metastable magnetic states. The coincidence between the Néel temperature and the onset of the resistance splitting is a strong evidence for the linking of the observed phenomena to the antiferromagnetism of the IrMn layer and not to the magnetic behavior of some Mn atoms dispersed in the tunneling barrier. We also note that the microscopic mechanism which yields the field-cool TAMR in IrMn is distinct from the high-field magnetotransport effects previously observed in iron pnictide AFMs.[9] In the latter materials the phenomenon has been ascribed to field-induced selection of structural crystal twin domains.[9] IrMn does not undergo a crystal phase change near $T_N$ and we therefore ascribe the distinct metastable states realized by field-cooling purely to distinct AFM configurations of uniform IrMn film.

The precise microscopic identification of these states requires a detailed study, which is beyond the proof-of-concept work presented in this paper. Here we recall the theoretical study[10] on IrMn which for $Ir_{20}Mn_{80}$ identified two non-collinear AFM phases 2Q and 3Q (confining the magnetic spins in the plane or yielding an out-of-plane component, respectively) with an energy difference of only ~0.25 mRy/atom, and a collinear phase whose energy is ~1.25 mRy/atom higher. We surmise that depending on the direction and strength of the applied field, the field-cooling procedure starting from temperatures above $T_N$ can favor spin configurations with different proportion of these distinct metastable AFM phases. Finally we remark that in the previously studied NiFe/IrMn exchange-spring AFM tunnel junctions,[2,3] the formation of the distinct magnetic configurations affecting the tunnel transport could be ascribed to bulk properties of the AFM or to the interface effects with the ferromagnet. From this perspective, our present experiments provide valuable complementary evidence showing that the interface with another magnetic layer is not required for stabilizing distinct states in the IrMn AFM.

To summarize, we have demonstrated the storage of information in an AFM/insulator/normal-metal tunneling device comprising no ferromagnetic elements. Different metastable configurations, yielding the high and low resistance states of the ATJ, can be set by cooling the AFM from above $T_N$ in magnetic fields with different orientations. By increasing the AFM layer thickness, the Néel temperature of the AFM film is expected to increase, virtually allowing setting the $T_N$ above room temperature. The absence of stray fields and the robustness against magnetic field perturbations are the key features of these devices, which hold potential for the development of novel spintronic devices without ferromagnets.

**Figures**

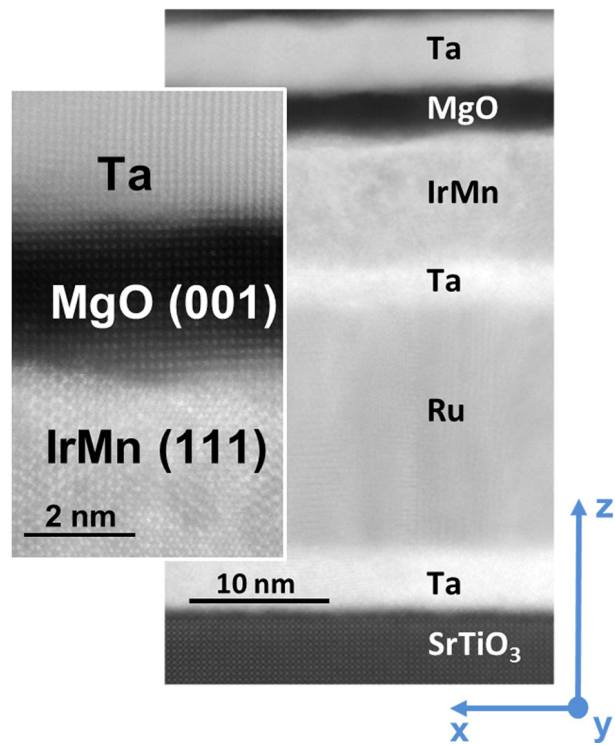

FIG. 1. (color online) High resolution Z contrast image of the heterostructure studied here. In the inset a high resolution image of the Ta/MgO/IrMn tunneling junction is shown. The reference system reported on the right is that used for indexing the magnetic fields during field cooling.

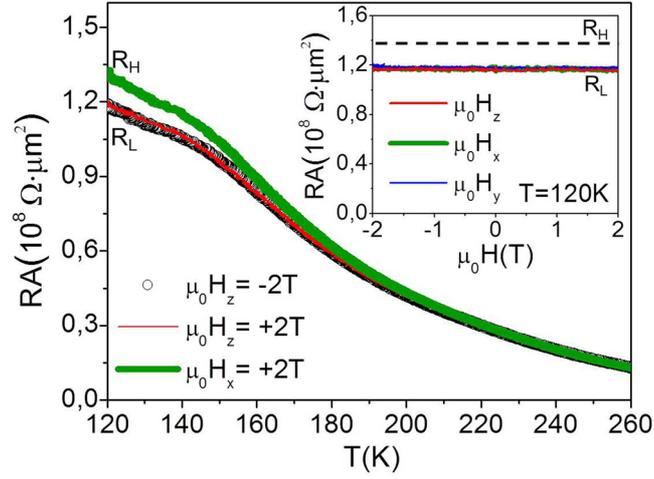

FIG. 2. (color online) Tunnel resistance data for field-cooling along positive and negative out-of-plane z-directions of the field and for the in-plane x-direction. The splitting of the two resistance traces, corresponding to the non-zero anisotropic magnetoresistance, is observed near $T_N$. Inset shows the stability of the state realized by field-cooling in the out-of-plane field. Below $T_N$, at T=120 K, the resistance remains constant when sweeping the magnetic field between +2 and -2T along out-of-plane (z) or orthogonal in-plane (x,y) directions.

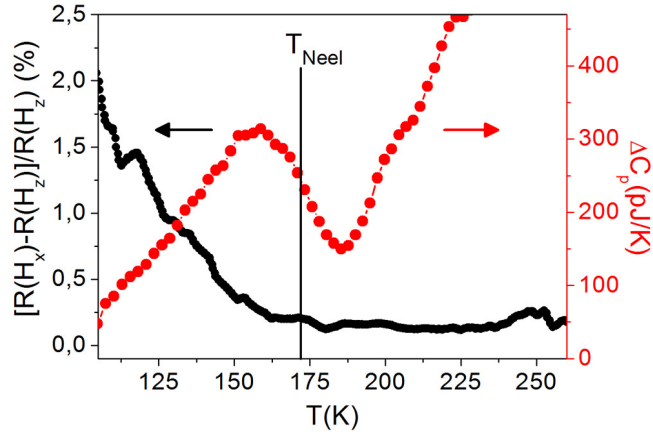

FIG. 3. (color online) Red scattered curve: Differential specific heat measurements of the 2 nm IrMn samples indicating $T_N$ ~173 K. Data were obtained on samples with a 2 nm thick IrMn layer, by averaging 1000 consecutive scans. Black curve: Temperature dependence of the tunneling magnetoresistance corresponding to the relative difference between field-cool resistance measurements in 2T fields applied along the out-of-plane (z) and in-plane (x) directions. The onset of a non-zero anisotropic magnetoresistance is observed when crossing $T_N$.

## Acknowledgements

We acknowledge fruitful discussions with G. Catalan, Jiun-Haw Chu, J.T. Heron and support from EU ERC Advanced Grant No. 268066, Ministry of Education of the Czech Republic Grant No. LM2011026, Academy of Sciences of the Czech Republic Preamium Academiae, Czech Science Foundation Grant P204/11/P339, and the JAE CSIC Grant. Research at ORNL supported by the U.S. Department of Energy (DOE), Basic Energy Sciences (BES), Materials Sciences and Engineering Division, and through a user project supported by ORNL's Shared Research Equipment (ShaRE) User Program, which is also sponsored by DOE-BES. Research at U. complutense supported by the ERC Starting Investigator Award "STEMOX". Research at Universitat Autonoma de Barcelona supported by a Marie Curie European Reintegration Grant within the 7th European Community Framework Programme, by the MICINN and by the Government of Catalonia through the Projects No. MAT2010-15202 and SGR2009-01225, respectively. D. Petti, E. Albisetti and R. Bertacco acknowledge financial support via the project FIRB "Ossidi nano strutturati: multifunzionalità e applicazioni" (RBAP115AYN).

**References**

[1] C. Chappert, A. Fert, and F. N. Van Dau, *Nat. Mater.* **2007**, *6*(11), 813-823.

[2] B. G. Park, J. Wunderlich, X. Marti, V. Holy, Y. Kurosaki, M. Yamada, H. Yamamoto, A. Nishide, J. Hayakawa, H. Takahashi, A. B. Shick, T. Jungwirth, *Nat. Mater.* **2011**, *10*, 347-351.

[3] X. Martí, B. G. Park, J. Wunderlich, H. Reichlová, Y. Kurosaki, M. Yamada, H. Yamamoto, A. Nishide, J. Hayakawa, H. Takahashi, and T. Jungwirth, *Phys. Rev. Lett.* **2012**, *108*, 017201 1-4.

[4] A. B. Shick, S. Khmelevskyi, O. N. Mryasov, J. Wunderlich, T. Jungwirth, *Phys. Rev. B* **2010**, *81*, 212409 1-4.

[5] T. Jungwirth, V. Novák, X. Marti, M. Cukr, F. Máca, A. B. Shick, J. Mašek, P. Horodyská, P. Němec, V. Holý, J. Zemek, P. Kužel, I. Němec, B. L. Gallagher, R. P. Campion, C. T. Foxon, J. Wunderlich, *Phys. Rev. B* **2011**, *83*, 035321 1-6.

[6] S A. Scholl, M. Liberati, E. Arenholz, H. Ohldag, and J. Stöhr, *Phys. Rev. Lett.* **2004**, *92*, 247201 1-4.

[7] Y. Y. Wang, C. Song, B. Cui, G. Y. Wang, F. Zeng, and F. Pan, *Phys. Rev. Lett.* **2012**, *109*, 137201 1-5.

[8] M. Molina-Ruiz, A. F. Lopeandía, F. Pi, D. Givord, O. Bourgeois, and J. Rodríguez-Viejo, *Phys. Rev. B* **2011**, *83*, 140407(R) 1-4.

[9] J.-H. Chu, J. G. Analytis, D. Press, K. De Greve, T. D. Ladd, Y. Yamamoto, and I. R. Fisher, *Phys. Rev. B* **2010**, *81*, 214502 1-.5

[10] A. Sakuma, K. Fukamichi, K. Sasao, and R. Y. Umetsu, *Phys. Rev. B* **2003**, *67*, 024420 1-7.